\documentclass[%
superscriptaddress,
nofootinbib,
twocolumn]{revtex4-1}

\usepackage{amssymb, amsmath, bm, dcolumn, epsf, graphicx, latexsym, slashed, simplewick}
\usepackage[utf8]{inputenc}

\usepackage{graphicx}
\usepackage{dcolumn}
\usepackage{bm}
\usepackage{xcolor}
\usepackage{bbold}
\usepackage{soul}
\usepackage[usestackEOL]{stackengine}
\usepackage{color}

\usepackage{amsmath}
\usepackage{amssymb}
\usepackage[normalem]{ulem}
\usepackage{float}

\usepackage{hyperref}

\usepackage{comment}

\begin{document}

\preprint{APS/123-QED}

\title{Domain Adaptation for Simulation-Based Dark Matter Searches\\Using Strong Gravitational Lensing}

\author{Stephon~Alexander}
\affiliation{Brown Theoretical Physics Center and Department of Physics, Brown University, Providence, RI 02912, USA}

\author{Sergei~Gleyzer}
\affiliation{Department of Physics \& Astronomy, University of Alabama, Tuscaloosa, AL 35401, USA}

\author{Pranath~Reddy}
\affiliation{Birla Institute of Technology \& Science, Pilani - Hyderabad Campus, Telangana, India}

\author{Marcos~Tidball}
\affiliation{Departamento de Física, Universidade Federal do Rio Grande do Sul, Porto Alegre, Brazil}

\author{Michael~W.~Toomey}
\affiliation{Brown Theoretical Physics Center and Department of Physics, Brown University, Providence, RI 02912, USA}

\date{\today}

\begin{abstract}
Clues to the identity of dark matter have remained surprisingly elusive, given the scope of experimental programs aimed at its identification. While terrestrial experiments may be able to nail down a model, an alternative, and equally promising, method is to identify dark matter based on astrophysical or cosmological signatures. A particularly sensitive approach is based on the unique signature of dark matter substructure on galaxy-galaxy strong lensing images. Machine learning applications have been explored in detail for extracting just this signal. With limited availability of high quality strong lensing data, these approaches have exclusively relied on simulations. Naturally, due to the differences with the real instrumental data, machine learning models trained on simulations are expected to lose accuracy when applied to real data. This is where domain adaptation can serve as a crucial bridge between simulations and real data applications. In this work, we demonstrate the power of domain adaptation techniques applied to strong gravitational lensing data with dark matter substructure. We show with simulated data sets of varying complexity, that domain adaptation can significantly mitigate the losses in the model performance. This technique can help domain experts build and apply better machine learning models for extracting useful information from strong gravitational lensing data expected from the upcoming surveys. 
\end{abstract}

\maketitle

 

\section{Introduction}

One of the great achievements of astrophysics in the last century was the realization by Zwicky, Rubin and others that the observed baryonic mass (stars, galaxies, etc.) was not consistent with the dynamics of galaxies and clusters. A natural solution to this problem was to consider some unseen matter that compensated for this discrepancy, or so-called dark matter. Presently, all efforts  aimed at extracting a non-gravitational signature of dark matter have come up empty \cite{Drukier:1986tm,Goodman:1984dc,Akerib:2016vxi,Cui:2017nnn,Aprile:2018dbl,2020arXiv200304545F,2015JCAP...09..008F,2015PhRvD..91h3535G,2018ApJ...853..154A,2017PhRvD..95h2001A,2020Galax...8...25R,2016JCAP...02..039M,2017arXiv170508103I,2015arXiv150304858T,2018PhRvL.120o1301D, 2015ARNPS..65..485G,2020arXiv200610735K,2020arXiv200612488B,Aaboud:2019yqu,2017JHEP...10..073S}. While this does not mean that dark matter can not communicate with Standard Model (SM) particles, as it may be the case that its SM couplings are strongly suppressed, there is also the possibility that such interactions do not exist.

Since its discovery, subsequent evidence for particle dark matter from its coupling to gravity is almost irrefutable \cite{planck,clust,wlens}. However, the list of possible models that fit current constraints is still quite broad. A particularly well suited signature that can be used to distinguish among dark matter models is the morphology and distribution of its substructure within dark matter halos. Some promising directions for inferring the properties of substructure include tidal streams \cite{Ngan:2013oga, 2016ApJ...820...45C, 2016PhRvL.116l1301B, 2016MNRAS.463..102E,Shih:2021kbt,Benito:2020avv} and astrometric observations \cite{Mishra-Sharma:2020ynk, VanTilburg:2018ykj, Feldmann:2013hqa, 2016arXiv160805624S,2020arXiv200811577V,Mishra-Sharma:2021nhh,Pardo:2021uzy}. A particularly sensitive probe will be strong gravitational lensing \cite{Buckley:2017ijx, Drlica-Wagner:2019xan, Simon:2019kmm} for which we restrict ourselves in this article.

Strong lensing has already seen some promising success in extracting information about dark matter substructure, from lensed quasars  \cite{sub1,sub2,sub3}, observations with ALMA \cite{alma}, and extended lensing images \cite{veg,koop,veko}. Various works have considered the expected signatures and methods to extract information about the underlying distribution of dark matter \cite{Daylan:2017kfh,2010MNRAS.408.1969V,pcat,subs}. 

More recently, there has been a plethora of applications of machine learning to this challenge, ranging from classification \cite{Alexander:2019puy, DiazRivero:2019hxf,Varma:2020kbq}, regression \cite{Brehmer:2019jyt}, segmentation analysis \cite{Ostdiek:2020mvo,Ostdiek:2020cqz}, and anomaly detection \cite{alexander_decoding_2020}. To date, all works have exclusively focused on the application of these techniques to simulations, in large part due to the limited availability of strong lensing data; something that is anticipated to change in the near future with the commissioning of the Vera C. Rubin Observatory and the launch of Euclid \cite{2019arXiv190205141V,2010MNRAS.405.2579O}. However, not unexpectedly, naively applying a model trained on simulations to real data is not likely to be very successful, as the data idiosyncrasies relative to simulations will significantly diminish the accuracy of the model. A promising method to bridge the gap between a model trained on simulations and real world data is based on the technique of domain adaptation (DA) \cite{36364}.

A subset of transfer learning, domain adaptation is focused on the generalization of the model across different domains or data sets drawn from different underlying distributions. Concretely, the goal of domain adaptation is to adapt a model trained on one data set (source) by generalizing it to another domain (target), where the objective of the model is unchanged. In practice domain adaptation can be realized in several ways, including supervised, semi-supervised, and unsupervised fashions \cite{2017arXiv170910190M,6618936,2020arXiv201003978F}.

Domain adaptation techniques have been used in a wide variety of applications related to computer vision, such as adapting a model trained on synthetic images to real images \cite{peng_visda_2017}, simple to complex images \cite{tzeng_adversarial_2017} and virtual worlds with controlled data to the real world \cite{schmidt_climategan_2021}. 

Recently, \cite{ciprijanovic_deepmerge_2021} used unsupervised domain adaptation (UDA) to classify merging galaxies. Doing so achieved promising results, with an increase of up to $19\%$ in accuracy when compared to a model trained only on simulations. This work also showed that models trained without domain adaptation, achieved a poor accuracy on real data. In another work \cite{Ostdiek:2020cqz} used domain adaptation to generalize an image segmentation algorithm to different gravitational lensing systems for subhalo detection.

In this work, we consider several domain adaptation algorithms to establish a proof-of-concept application for dark matter searches in strong gravitational lensing. Given the present lack of sufficient real data, we use two datasets with differing levels of complexity of strong lensing simulations to carefully test the performance of domain adaptation prior to eventual applications to real data. We evaluate the application of models trained on the source data set to identify various types of dark matter substructure on the target data set. We compare the performance of several domain adaptation algorithms, to find the best model. We additionally investigate equivariant neural network models that incorporate a known group symmetry, to further enhance the performance of domain adaptation.

We begin with a brief review of dark matter substructure and also discuss several lensing signatures in Section \ref{sec:DMS}. We then focus on the details of strong lensing simulations in Section \ref{sec:SLS}, followed by a summary of domain adaptation algorithms in Section \ref{sec:ALG}. We present our main results in Section \ref{sec:RES}, followed by the discussion and outlook in Section \ref{sec:DNC}.

\section{Dark Matter Detection and Strong Gravitational Lensing}\label{sec:DMS}

\subsection{Dark Matter Substructure}

The $\Lambda$ Cold Dark Matter ($\Lambda$CDM) model envisions near scale invariant density fluctuations, present in the early universe, serving as seeds of large-scale structure via hierarchical structure formation. Concretely, structures such as dark matter halos are formed from the coalescence of smaller halos \cite{Kauffmann:1993gv}. Evidence for such merges has been observed in our Galaxy \cite{2021NatAs...5..392C,Necib:2019zka,Necib:2019zbk} and is a general prediction of N-body simulations where evidence of mergers should remain largely in tact. The distribution of subhalos masses is expected to follow a power law distribution,
\begin{equation}
    \frac{d N}{d m} \propto m^\beta,
    \label{subdist}
\end{equation}
where $\beta \sim -1.9$ has been found from simulations \cite{Springel:2008cc, Madau:2008fr}. 

Comparison between simulation and observation indicates good agreement with $\Lambda$CDM on large-scales \cite{planck,clust,wlens}. However, discrepancies begin to arise on smaller, sub-galactic scales. These include the core-vs-cusp \cite{burk,Oh_2015}, too big to fail \cite{Boylan_Kolchin_2011}, missing satellite \cite{Moore_1999,Klypin_1999,msp},\footnote{See \cite{mspr} for a differing perspective.} and diversity problems \cite{Oh_2015}. While it may be the case that these problems can be addressed with a better understanding of the astrophysics, e.g. baryonic feedback \cite{2019MNRAS.488.2387B}, it is imperative that we consider the manifestations of other theories beyond $\Lambda$CDM.

\begin{figure*}[t]
    \centering
    \includegraphics[width=0.377\linewidth]{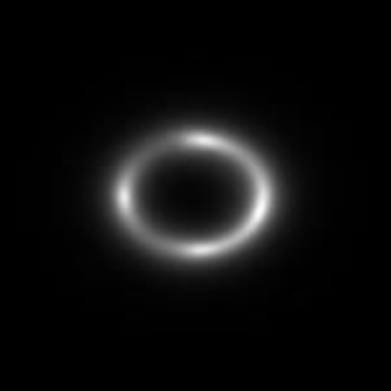}\hspace{1cm}
    \includegraphics[width=0.38\linewidth]{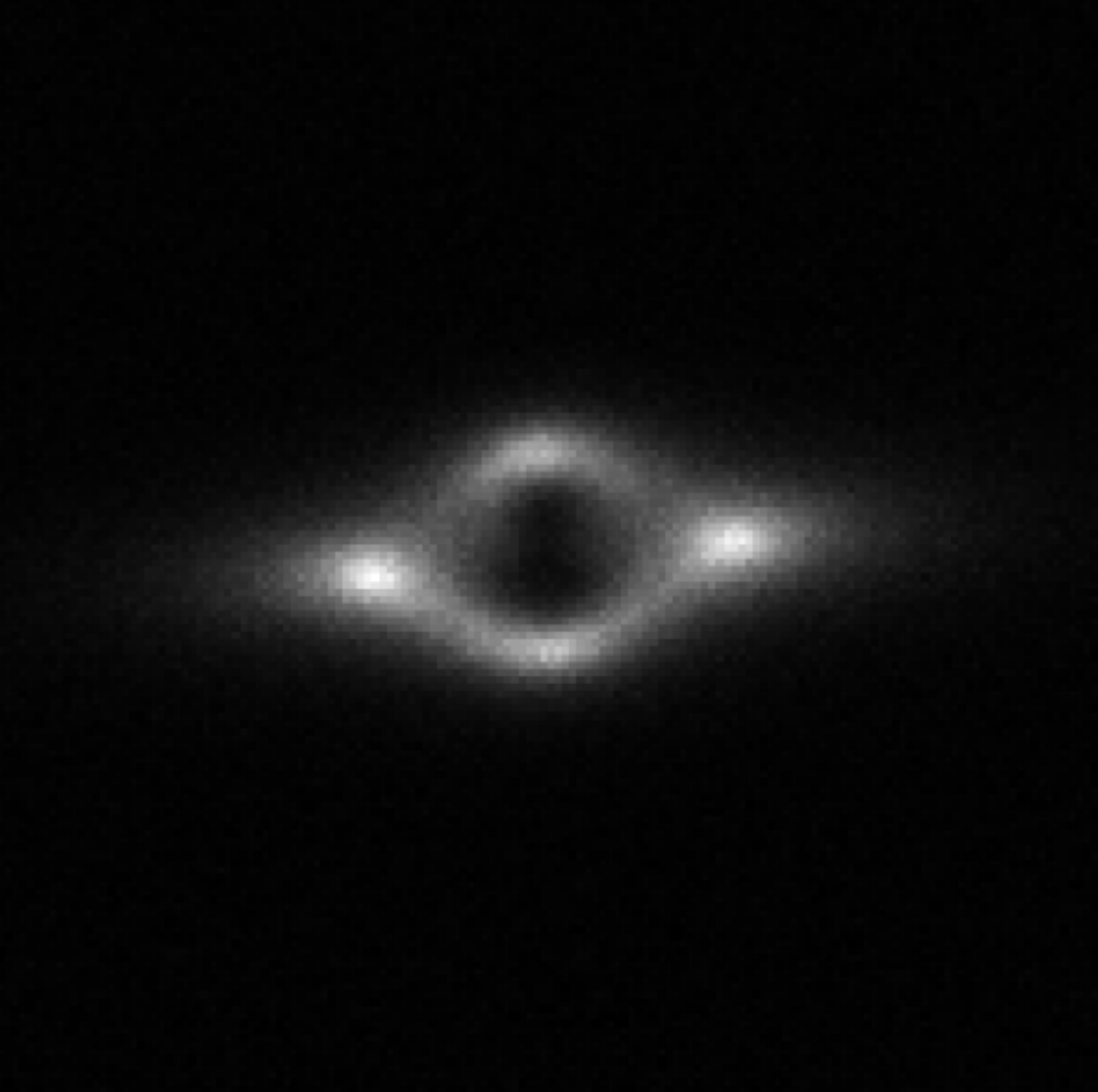}
    \caption{Example images from the source (left) and target (right) data set.}
    \label{fig:source_target_example}
\end{figure*}

Two natural directions to consider are a modification to the general theory of relativity, such as BF coupled \cite{Alexander:2018tyf,Alexander:2020tsf,Alexander:2021uny} or Chern-Simons gravity \cite{Jackiw:2003pm,Alexander:2009tp}, or alternatives to cold non-interactive dark matter. In the context of this paper, we will focus on the latter. An example of a dark matter model that addresses several of the tensions mentioned above is condensate dark matter, which can be realized in the context of Bose-Einstein (BEC) \cite{Sin:1992bg,Silverman:2002qx,Hu:2000ke,Sikivie:2009qn,Hui:2016ltb,Berezhiani:2015bqa,Ferreira:2018wup} or Bardeen-Cooper-Schreifer(BCS) \cite{Alexander:2016glq,Alexander:2018fjp,Alexander:2020wpm} condensates. A concrete and well motivated model is the axion. As the Goldstone boson of a broken U(1) symmetry, axions were originally introduced as a solution to the Strong-CP problem \cite{Peccei:1977hh,Wilczek:1977pj,Weinberg:1977ma}. Shortly thereafter, it was recognized that they were a promising dark matter candidate \cite{Preskill:1982cy,Abbott:1982af,Dine:1982ah}. Very light axions are particularly well suited to address some of the issues with structure on sub-galactic scales. Axions with masses $\sim 1 \times 10^{-23}$~eV have a de~Broglie wavelength on kpc scales, which realizes a natural solution to the core-vs-cusp problem. Additionally, light axions can form subhalos but can also form substructures quite different from standard CDM. These include vortices, disks, and interference patterns \cite{sfdm,Hui:2021tkt,Hui:2020hbq,Alexander:2019qsh,Alexander:2021zhx}.

While we will not consider their impact in this work, it is important to also consider line of sight halos, i.e. interlopers \cite{Sengul:2020yya,McCully_2017,Despali_2018,Gilman_2019}. In some cases their influence may dominate the signal of substructure, so it is important to take care that they are not incorrectly associated with the dark matter halo when working with real data sets (see for example \cite{Sengul:2021lxe}). 

\subsection{Strong Lensing Theory}

A powerful probe of dark matter substructure is strong gravitational lensing; an effect which is most pronounced near extended lensing arcs. Any over or under densities along the line of sight for an observer yields the \textit{lens equation}, realized as an integral over an induced gravitational potential \cite{Nar-Bart:1997lens}, 
\begin{equation}
    \vec{\beta} = \vec{\theta} - \frac{2}{c^2} \frac{D_{LS}}{D_S D_L}\vec{\nabla}\int dz ~ \Psi(\vec{r}),
    \label{eq:lens_eq}
\end{equation}
where $D_{LS}, D_L, D_S$ are the angular diameter distances from the lens to the source, the lens to the observer, and the source to observer. The last term on the r.h.s. of Equation \ref{eq:lens_eq} is known as the deflection angle, $\vec{\alpha}$, and is related via a perpendicular derivative to the the lensing potential, $\Psi_{;\perp} = \alpha$. The lensing potential can be shown to be related to matter density via a Poisson equation for gravitational lensing,
\begin{equation}
    \nabla^2 \Psi = 2 \kappa,\label{poiss}
\end{equation}
implying that lensing from separate sources, a DM halo and its subhalos, is a sum of deflection angles,
\begin{equation}
    \vec{\alpha} = \sum_{\rm all~matter} \vec{\alpha}_i,
\end{equation}
where $\vec{\alpha}_i$ could represent a DM halo, subhalos, vortices, external shear, interlopers etc. 

\section{Strong Lensing Simulations} \label{sec:SLS}

Similar to \cite{Alexander:2019puy,alexander_decoding_2020} we consider data sets of three substructure classes; no substructure, subhalos of CDM, and vortex substructure of superfluid type dark matter. The parameters for the simulations are complied in  Table \ref{tab:table}. The strong lensing images are simulated with \texttt{PyAutolens} \cite{2015MNRAS.452.2940N,2018MNRAS.478.4738N}. The data are sized $150 \times 150$ pixels with a scale $0.5''/$pixel. The light from lensed background galaxies is modeled as a simple Sersic profile and the signal-to-noise ratio (SNR) of the lensing arcs are consistent with real lensing data -- $SNR \sim 20$ \cite{hst} -- by appropriate inclusion of backgrounds and noise. We further consider the modifications induced by a point-spread function (PSF), modeled as an Airy disk with first zero-crossing $\sigma_{\rm psf} \lesssim 1''$.

We ensure that the total fraction of mass in substructure, $f_{sub}$,  is of $\mathcal{O}(1\%)$. We constrain the simulations such that the total mass of the halo, including substructure, is always equal to $1 \times 10^{12} ~ {\rm M}_\odot$. This is done to ensure that classification algorithms don't simply recognize simulations without substructure as less massive on average. When simulating substructure for CDM we draw subhalo masses from from Equation \ref{subdist} for a total number of sources taken from a Poisson draw with $\mu = 25$, consistent with the expected number of subhalos for our field of view and redshift range \cite{Rivero:2018bcd}. We model vortices of superfluid dark matter as uniform density strings of mass of varying length and orientation -- a valid approximation at cosmological distances. Beyond the effects of substructure, we also include the impact of external shear due to large-scale structure. 

The inclusion of the effects of substructure in our simulations can be understood from the linearity of the Poisson equation, Equation \ref{poiss}, which implies the total lensing is just the sum of the individual contributions,
\begin{equation}
    \alpha = \alpha_{LSS}  + \alpha_{halo} + \alpha_{halo-sub},
\end{equation}
where $\alpha_{LSS}$ is the external shear from large-scale-structure, $~\alpha_{halo}$ the lensing from the halo and $~\alpha_{halo-sub}$ for halo substructure. It then follows that the location of an image can be found from a modified form of the lens equation, Equation \ref{eq:lens_eq},
\begin{equation}
    \beta^i = \theta^i - \alpha^i_{LSS}  - \alpha^i_{halo} - \alpha^i_{halo-sub}.
\end{equation}

Domain adaptation requires at least two data sets, the source and the target. In this work we consider two distinct simulations as source and target data sets. The biggest difference between source and target data sets is that source images are held at fixed redshift, while the target data set allows the redshifts to float over a range of  values for both the lensed and lensing galaxy. Additionally, the SNR in the target data set varies over a larger range of values, $ 10 \lesssim {\rm SNR}\lesssim 30$, consistent with the challenges associated with variable source distances. Thus, the target data set constitutes a more challenging sample for identifying dark matter substructure - this difference is easy to see by eye in Figure \ref{fig:source_target_example} for example lenses from both classes. Thus, it will be our goal to successfully adapt and evaluate the algorithms trained on the source data set to the target data set. 

\begin{table}
    \scriptsize
	\centering
	\caption{Parameters with distributions and priors used in the simulation of strong lensing images. Parameters with subscript $s$ \& $t$ correspond to parameters for source and target data sets respectively. Note that only a single type of substructure was used per image.}
	\vspace{0.2cm}
	\label{tab:table}
	\begin{tabular}{cccc}
		\hline
		\hline
		DM Halo\\ 
		\hline
		\hline
		\textbf{Param.} & \textbf{Dist.} & \textbf{Priors} & \textbf{Details} \\
		\hline
$\theta_x$  &  fixed & 0 & x position \\
$\theta_y$  &  fixed & 0 & y position \\
z$_s$  & fixed & 0.5  & redshift\\
z$_t$  & uniform & [0.4,0.6]  & redshift\\
$M_{\rm TOT}$ & fixed & 1e12 & total halo mass in $ {\rm M}_\odot$\\
        \hline
		\hline
		Ext. Shear\\
		\hline
		\hline
		\textbf{Param.} & \textbf{Dist.} & \textbf{Priors} & \textbf{Details} \\
		\hline
		$\gamma_{ext}$ & uniform & [0.0, 0.3] & magnitude \\
		$\phi_{ext}$ & uniform & [0, 2$\pi$] & angle \\
        \hline
		\hline
		Lensed Gal.\\ 
		\hline
		\hline
		\textbf{Param.} & \textbf{Dist.} & \textbf{Priors} & \textbf{Details} \\
		\hline
$r$ & uniform & [0, 0.5] & radial distance from center\\
$\phi_{bk}$ & uniform & [0, 2$\pi$] & orientation from y axis\\ 
z$_s$  &  fixed & 1.0  & redshift\\
z$_t$  &  uniform & [0.8,1.2]  & redshift\\
e  &  uniform & [0.4, 1.0] & axis ratio\\
$\phi$ & uniform & [0, 2$\pi$] & orientation to y axis \\

n & fixed & 1.5 & Sersic index \\
R & uniform & [0.25,1] & effective radius \\
        \hline
		\hline
		Vortex \\
		\hline
		\hline
		\textbf{Param.} & \textbf{Dist.} & \textbf{Priors} & \textbf{Details} \\
		\hline
$\theta_x$  &  normal & $[0.0,0.5]$ & x position \\
$\theta_y$  & normal & $[0.0,0.5]$ & y position\\
$l$ & uniform & [0.5,2.0] & length of vortex\\
$\phi_{vort}$ & uniform & [0, 2$\pi$] & orientation from y axis\\ 
$m_{vort}$ & uniform & [3.0,5.5] & \% of mass in vortex\\
        \hline
		\hline
		Subhalo \\
		\hline
		\hline
		\textbf{Param.} & \textbf{Dist.} & \textbf{Priors} & \textbf{Details} \\
		\hline
$r$ & uniform & [0, 2.0] & radial distance from center\\
$N$ & Poisson & $\mu$=25 & number of subhalos\\
$\phi_{sh}$ & uniform & [0, 2$\pi$] & orientation from y axis\\ 
$m_{sh}$ & power law & [1e6,1e10] & subhalo mass in ${\rm M}_{\odot}$\\
$\beta_{sh}$ & fixed & -1.9 & power law index

	\end{tabular}
\end{table}

\section{Domain Adaptation} \label{sec:ALG}

Our goal is to train a supervised model on a source data set and adapt it to a target data set. For this task we use \textit{ResNet-18} \cite{he_deep_2015}, a Convolutional Neural Network (CNN), as our base architecture. This is the same architecture that has achieved top performance in previous applications to lensing data sets \cite{Alexander:2019puy,alexander_decoding_2020,2020arXiv200811577V}. More generally, CNNs are known to outperform other methods of classification for strong gravitational lenses \cite{metcalf_strong_2019}, nonetheless, as noted by \cite{ciprijanovic_deepmerge_2021}, a model trained on simulations can perform poorly on real data.

To improve the performance of models trained on simulated data, we use unsupervised domain adaptation, which attempts to mitigate the effects of the domain shift between the source and the target domains. It enables a transfer of knowledge gained from a labeled source data set to a distinct unlabeled target data set, within the constraint that the objective remains the same \cite{french_self-ensembling_2018}. Example source and target data are shown in Fig. \ref{fig:source_target_example}. For our analysis we consider three UDA algorithms described below.

\begin{figure}[t]
    \centering
    \includegraphics[width=1\linewidth]{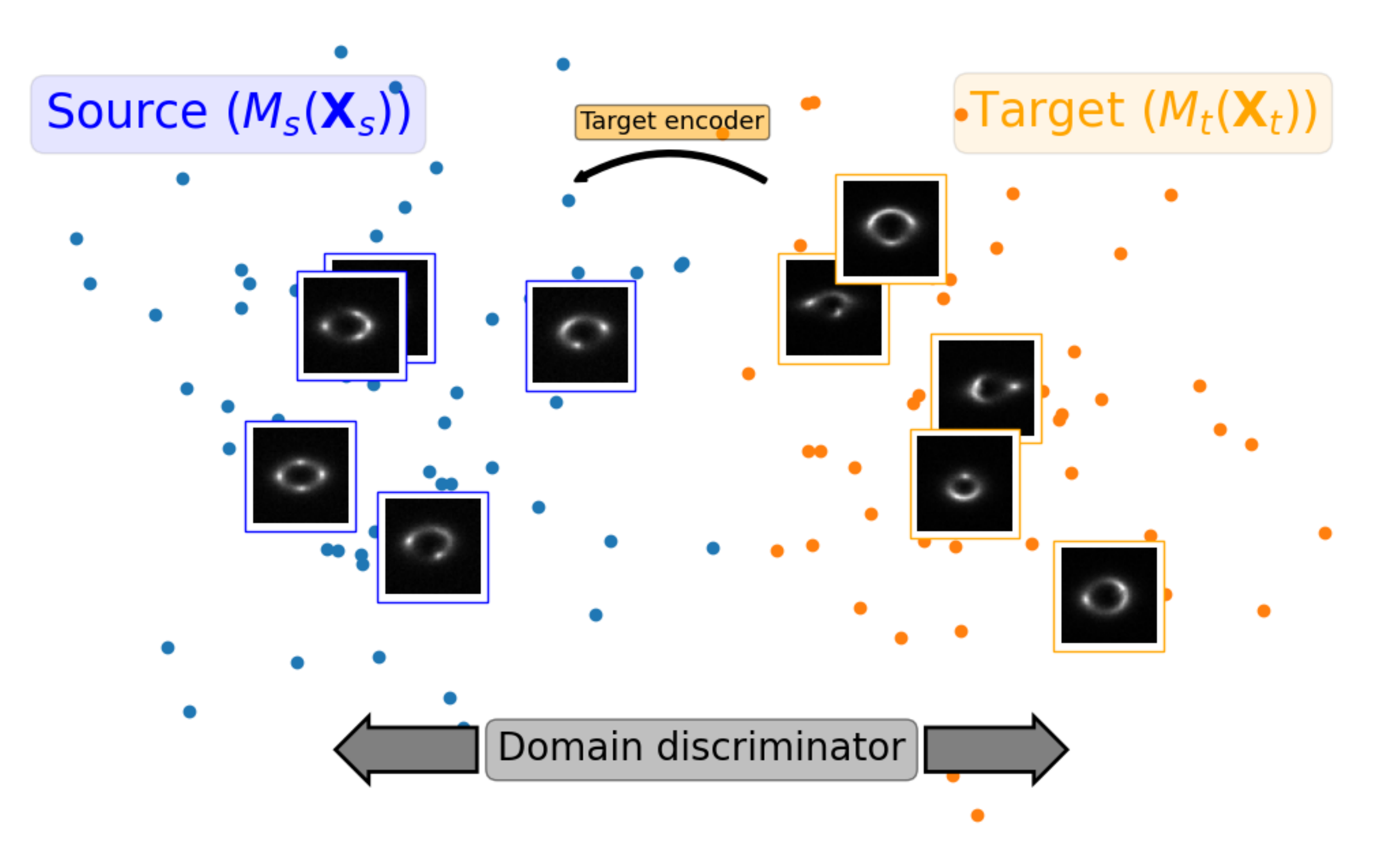}
    \caption{Depiction of target and source representations in the latent space of the encoder in ADDA. Images in blue are from the source domain and images in orange are from the target domain.}
    \label{fig:adda}
\end{figure}

\begin{table*}
	\centering
	\caption{Hyperparameters and augmentations used to train the UDA algorithms.}
	\vspace{0.2cm}
	\label{tab:hparams}
			\bgroup
\def\arraystretch{1.25}
	\begin{tabular}{|l|l|c|l|}
		\hline
		\textbf{Method} & \textbf{Hyperparameters} & \textbf{Hyperparameter values} & \textbf{Augmentations}\\
		\hline
 & Learning rate & $1\times10^{-3}$ & Horizontal flips \\
Supervised & Weight decay & $5\times10^{-5}$ & Vertical flips \\
  & Cyclic scheduler & True &  \\
\hline
 & Learning rate target encoder& $1\times10^{-5}$ & Horizontal flips \\
ADDA & Learning rate target encoder & $1\times10^{-4}$ & Vertical flips \\
  & Weight decay & $5\times10^{-5}$ &  \\
  & Cyclic scheduler & False &  \\
\hline
 & Learning rate& $1\times10^{-3}$ & Horizontal flips \\
AdaMatch  & Tau & $0.9$ & Vertical flips \\
  & Weight decay & $5\times10^{-5}$ & Random autocontrast \\
  & Cyclic scheduler & False & Gaussian blur \\
  & & & Random invert \\
  & & & Random adjust sharpness \\
  & & & Random solarize \\
  & & & Random affine transformations \\
\hline
  & Learning rate& $1\times10^{-3}$ & Horizontal flips \\
Self-Ensemble  & Weight of unsupervised loss & $3.0$ & Vertical flips \\
  & Weight decay & $5\times10^{-5}$ & Gaussian blur \\
  & Cyclic scheduler & False &  \\
\hline

	\end{tabular}
	\egroup
\end{table*}

\subsection{Unsupervised Domain Adaptation Techniques}

We first consider Adversarial Discriminative Domain Adaptation (ADDA) \cite{tzeng_adversarial_2017}, an adversarial adaptation method with the goal of minimizing the domain discrepancy distance through an adversarial objective with respect to a discriminator. Ideally the discriminator will be unable to distinguish between the source and the target distributions. We consider that we have access to source images $\mathbf{X}_s$ and labels $\mathbf{Y}_s$ that come from a source distribution $p_s (x, y)$ and also target images $\mathbf{X}_t$ from a target distribution $p_t(x,y)$. Our objective is to learn a target encoder $M_t$ and classifier $C_t$ that classifies $\mathbf{X}_t$ into $K$ classes.

Due to the fact that it is not possible to perform supervised learning on the target distribution, we learn a source encoder $M_s$ and a source classifier $C_s$. The encoder learns to map the input samples to a latent vector whose dimensionality is lower than the dimensionality of the input samples. With these networks trained, the distance between $M_s(\mathbf{X}_s)$ and $M_t(\mathbf{X}_t)$ is minimized, as illustrated in Fig. \ref{fig:adda}. Since we're only minimizing the encoders, we can assume that $C_s = C_t = C$.

We train $M_s$ and $C$ using a standard supervised loss. Then, we train a discriminator $D$ that classifies if the encoded vector represents an image from the source domain or from the target domain using a standard supervised loss, where the labels indicate the origin domain. Finally, we train the $M_t$ using $D$. To evaluate a target image $\mathbf{X}_t$ we perform $C(M_t(\mathbf{X}_t))$.

We additionally evaluate two DA algorithms derived from semi-supervised learning. The first makes use of self-ensembling (Self-Ensemble) \cite{french_self-ensembling_2018} and is based on the mean teacher semi-supervised model \cite{2017arXiv170301780T}. There are two networks in this method: a student network that is trained using gradient descent and a teacher network whose weights are an exponential moving average of the student's. During training, the labeled source inputs are passed through the student network and the cross-entropy loss is taken. However, the unlabeled target inputs pass through both the student and teacher networks and the self-ensembling loss is used. It is computed as the mean-squared difference between the predictions created by the student and the teacher networks with different augmentations, dropout and noise parameters, and penalizes the difference in class prediction between the student and the teacher. This method also makes use of confidence thresholding and a class balancing loss term. The second semi-supervised method, AdaMatch \cite{berthelot_adamatch_2021}, uses weak and strong augmentations on both the source and target input images. During training the method also uses random logit interpolation, distribution alignment and relative confidence thresholding to achieve a better performance.

In additional to the baseline CNN models, we also consider an Equivariant Neural Network (ENN) \cite{weiler_general_2021} for substructure classification. ENNs can be thought of a generalization of a CNN that encode the representation of a useful symmetry, both global or local, such that its group convolutions are invariant symmetries present in the data. This is useful if there is a known symmetry in the problem. As we expect lensing images to have symmetries beyond simple translation, for example rotations, the flexibility of choosing different group representations is expected to improve the performance. 

\begin{figure}
    \centering
    \includegraphics[width=\linewidth]{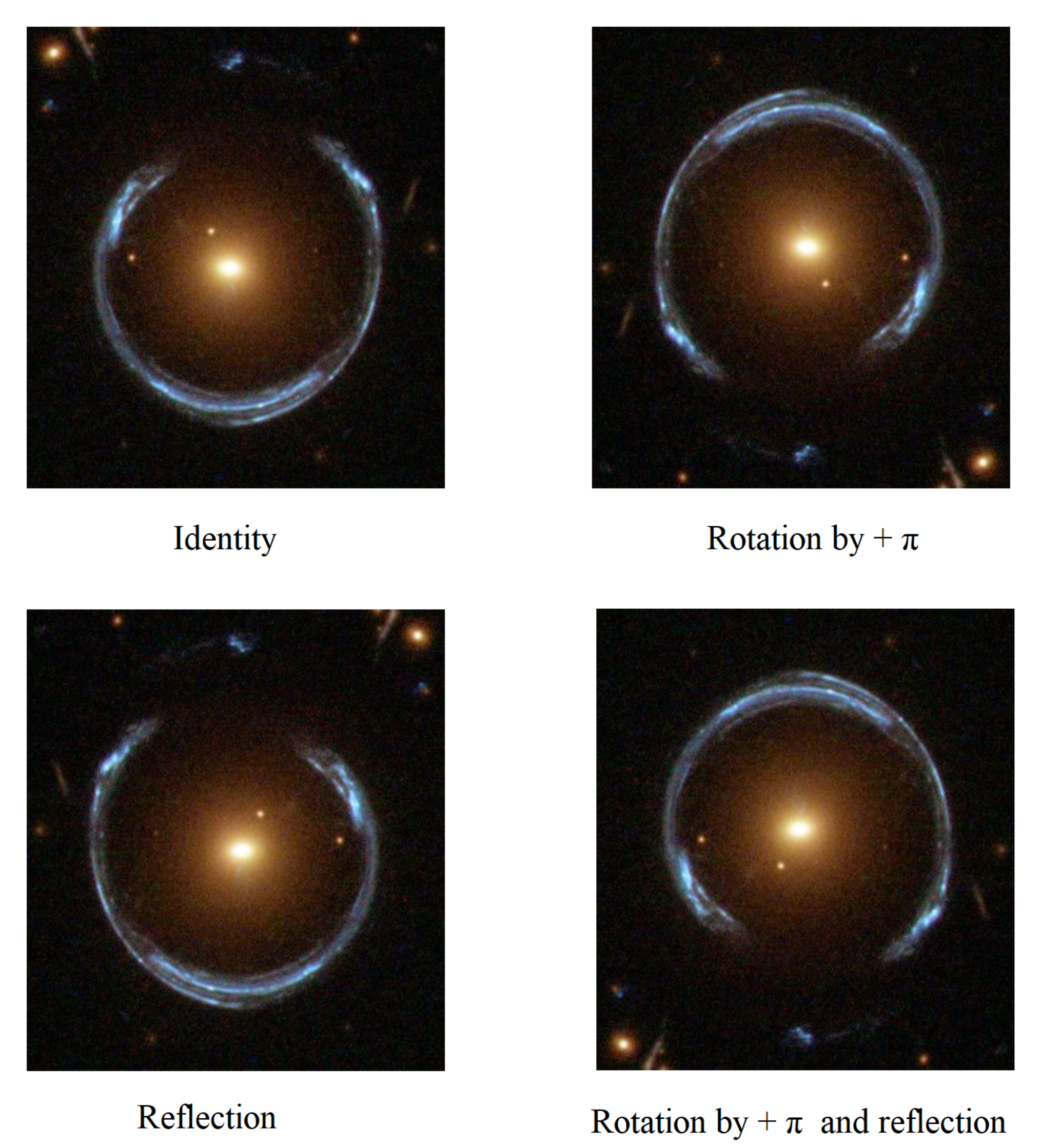}
    \caption{Visualization of the $D_2$ group structure for a real gravitational lens. Lensing image credit: ESA/Hubble \& NASA.}
    \label{fig:group}
\end{figure}

\begin{figure}
    \centering
    \includegraphics[width=0.95\linewidth]{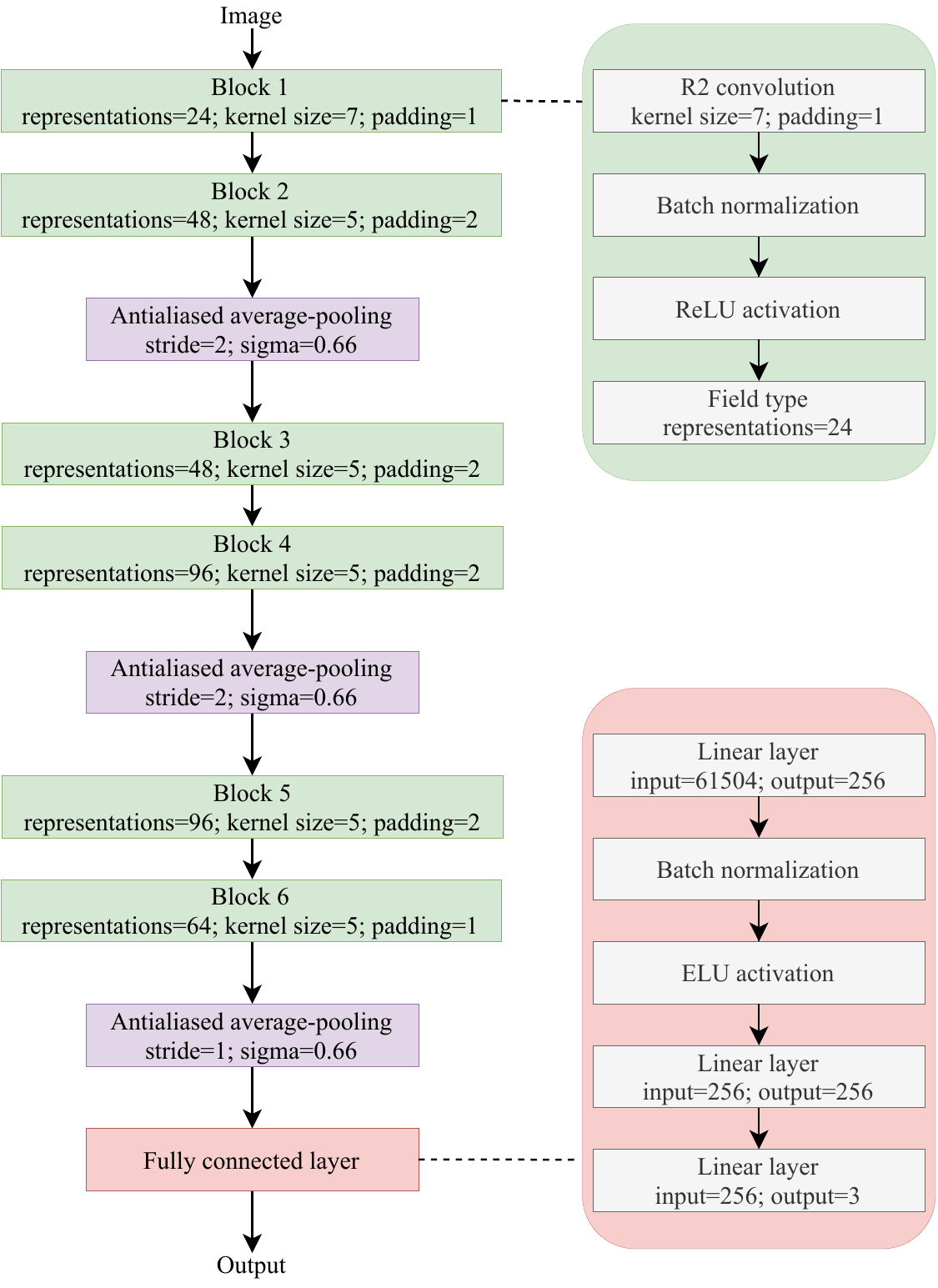}
    \caption{Schematic of the equivariant (ENN) architecture.}
    \label{fig:enn}
\end{figure}

The ENN we use consists of a group equivariant convolutional neural network \cite{cohen2016group} with six equivariant convolution blocks. We utilize the dihedral group $D_2$, whose symmetry mappings include the identity, rotations by $\pm \pi$ and horizontal/vertical reflections. The $D_2$ group structure can be visualized in Fig. \ref{fig:group}. Each block is composed of a convolutional layer, a batch normalization layer and a ReLU activation function. After each pair of layers we perform channel-wise average-pooling and in the end we use a fully connected layer for multiclass classification. A schematic of the ENN architecture is presented in Fig.~\ref{fig:enn}.

\subsection{Network Training}

For training we use $30,000$ images for the source domain and $30,000$ images for the target domain; in both cases there are $10,000$ images per class. For validation we use $7,500$ images for the source domain and $7,500$ images for the target; in both cases there are $2,500$ images per class. We used the \textit{Adam} optimizer \cite{kingma_adam_2017} to minimize our losses. We trained both \textit{ResNet-18} and the ENN for $200$ epochs, training with a patience of $15$ epochs, such that if the accuracy of the model does not improve in $15$ epochs we stop training. For final results, we considered the epoch that achieved the largest accuracy. Learning rate, weight decay and other hyperparameters were optimized through a hyperparameters search, and are available in Table \ref{tab:hparams}.

We used random horizontal and random flips augmentations for both the source and target dataset. We also found that the best results were obtained after random zooming (in a range of $[0.8, 1.2]$) and random rotations (in a range of $[0, 90]$ degrees) on the source dataset. Two of the UDA algorithms, Self-Ensemble and AdaMatch, are also highly dependent on augmentations, and different augmentations were tested to find the optimal configurations. We utilize the area under the ROC curve (AUC) on the target validation set as the metric for classifier performance for all the models. All quoted AUC values are macro-averaged unless stated otherwise. All machine learning models were implemented using \texttt{PyTorch} \cite{paszke_pytorch_2019} and are run on a single NVIDIA Tesla P100 GPU. 

\section{Results} \label{sec:RES}

\begin{figure*}[!t]
    \centering
    \includegraphics[width=0.85\linewidth]{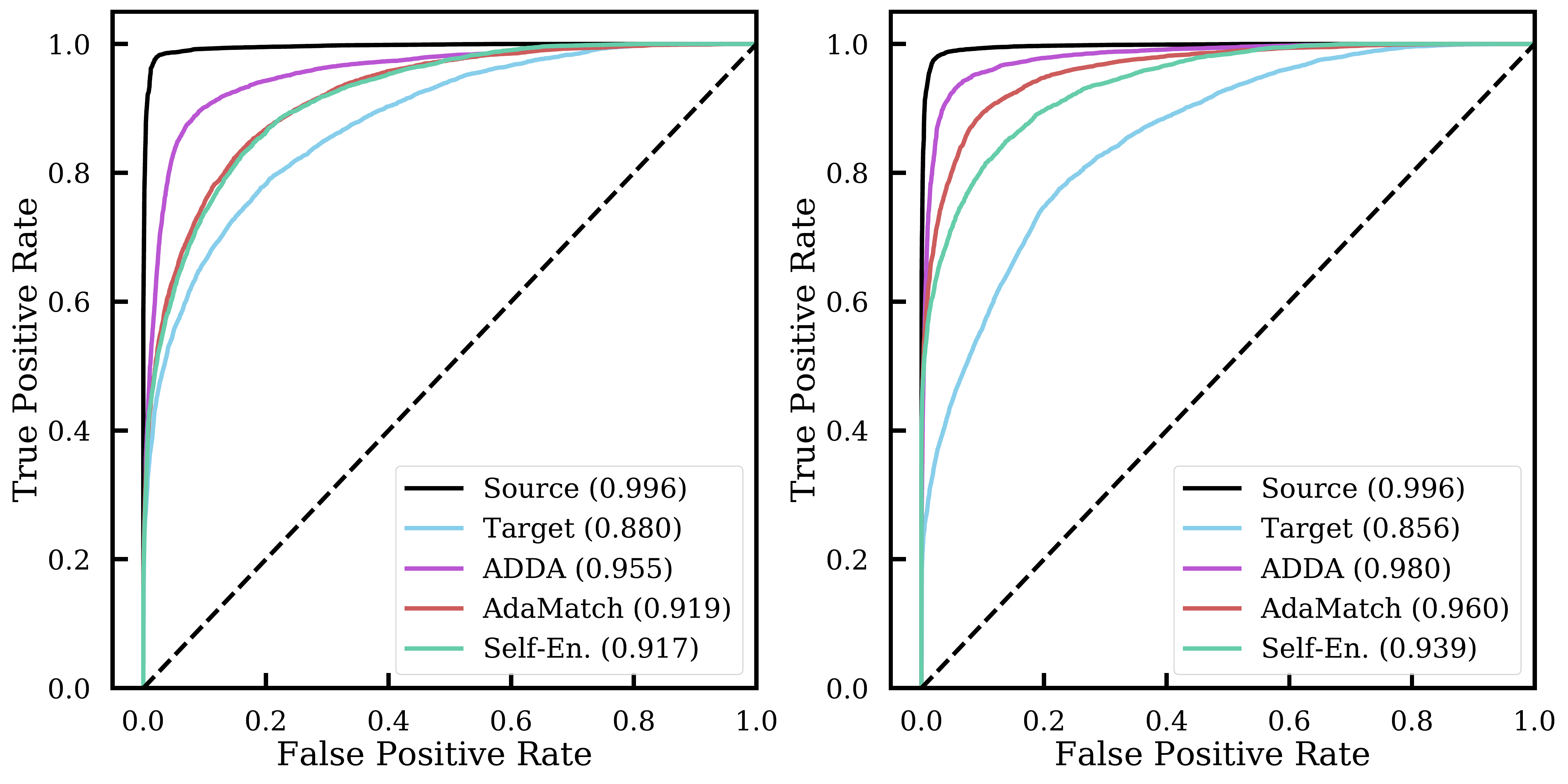}
    \caption{ ROC-AUC curves for \textit{ResNet-18} (left) and ENN (right) based algorithms applied to our data set.}
    \label{fig:roc}
\end{figure*}

We compare three different UDA techniques in the context of multi-class classification of three types of substructure: no substructure, CDM subhalos, and superfluid DM vortices. We employ two different base classifiers for each technique - \textit{ResNet-18} and an ENN.\footnote{The code used in our analysis can be found \href{https://github.com/ML4SCI/DeepLense/tree/main/Domain_Adaptation_for_DeepLense_Marcos_Tidball}{here}. } The parameters for our source and target data sets are shown in Table \ref{tab:table}. Better results on all methods were achieved when starting UDA training with models that were pre-trained in a supervised fashion using source data, as discussed in \cite{ciprijanovic_deepmerge_2021}. As such, our models were trained starting from models pre-trained on the source simulation data. Results from our analysis are shown in Tables \ref{tab:resnet} and \ref{tab:enn} for \textit{ResNet-18} and the ENN respectively. ROC curves for both architectures are presented in Figure~\ref{fig:roc}.

\subsection{Domain Adaptation}

We first train \textit{ResNet-18} on the source data set where it achieved a macro-averaged AUC $\approx~0.996$ and an accuracy of $\approx$ 97\%. Applying this model to the target data set naively, i.e. \textit{without} domain adaptation, results in an AUC of $\approx$ $0.880$ and accuracy of $\approx$ 59\%, a significantly degraded performance, as anticipated. 

Following the application of UDA techniques, we observe a significant improvement when applied to the target data set. AdaMatch and Self-Ensemble achieve an AUC of $\approx$ $0.919$ and $0.917$, respectively, which shows good improvement over the results obtained without domain adaptation. The top performing \textit{ResNet-18} based UDA model is ADDA. With a $\approx$ 9\% improvement in AUC, at $\approx$ $0.955$, and accuracy of $\approx$ 86\%, ADDA significantly improves the performance of our model on the target data set. 

Considering the performance for individual classes, we find that ADDA achieves consistent AUC scores of $\approx 0.95$ for all three classes. This is in contrast to the performance on naive application, where no substructure and subhalos classification each had an AUC of $\approx 0.90$ and classification of vortices was severely degraded to an AUC of $\approx 0.80$. The baseline model applied to the source data set, on the other hand, had consistent performance on a class by class basis at $\approx 0.99$ AUC.

\begin{table}[!h]
	\centering
	\caption{Accuracy and macro-averaged area under the ROC curve (AUC) for the \textit{ResNet-18} classifier.}
	\vspace{0.2cm}
	\label{tab:resnet}
		\bgroup
\def\arraystretch{1.25}
	\begin{tabular}{|ccc|c|ccc|}
		\hline
		&\textbf{Method} & &~\textbf{Accuracy (\%)} & &\textbf{AUC}&\\
		\hline
~~ & ADDA &~~~& $85.84$ &~~~& $0.955$ & ~~~\\
&AdaMatch & &  $75.55$ && $0.919$ &\\
&Self-Ensemble & &  $76.71$ && $0.917$ & \\
&Supervised (target) & &  $59.19$  &&  $0.880$ & \\ 
&Supervised (source) & &  $96.84$  &&  $0.996$ & \\ \hline
	\end{tabular} 
	\egroup
\end{table}

\subsection{Equivariant Domain Adaptation}

With an AUC of $\approx$ 0.996 and accuracy of $\approx 91 \%$, the ENN performance is commensurate with  \textit{ResNet-18} on the source data set. The naive application to the test data set (i.e. no domain adaption) again results in degraded performance, realized with an AUC $\approx$ 0.856 and accuracy of $\approx$ 68\%. 

The Self-Ensemble and AdaMatch equivariant models again show improved performance relative to the naive case and with AUC scores of $\approx$ 0.939 and $\approx$ 0.960 respectively show a notable performance boost relative to the same UDA algorithms with \textit{ResNet-18}. This is easily visualized by comparing the increased separation between the target ROC curve (blue) with the three UDA algorithms in Figure \ref{fig:roc} between \textit{ResNet-18} (left) and the ENN (right). Finally, ADDA again achieves impressive top performance with an AUC of $\approx$ 0.980. Thus, the equivariant model with ADDA is the top performing algorithm relative to the baseline source model.

\begin{table}[!h] 
	\centering
	\caption{Accuracy and macro-averaged area under the ROC curve (AUC) for the ENN classifier.}
	\label{tab:enn}
	\vspace{0.2cm}
	\bgroup
\def\arraystretch{1.25}
	\begin{tabular}{|ccc|c|ccc|}
		\hline
		&\textbf{Method} & & ~\textbf{Accuracy (\%)} & &
		\textbf{AUC} & \\ \hline
~~& ADDA  &~~~&  $91.47$ &~~~& $0.980$ &~~~\\
&AdaMatch  &~~~& $85.81$ &~~~& $0.960$ & \\
&Self-Ensemble  &~~~& $80.09$ &~~~& $0.939$ &\\
&Supervised (target)  &~~~& $67.53$  &~~~&  $0.856$ & \\ 
&Supervised (source)  &~~~& $97.09$  &~~~& $0.996$ & \\ \hline
	\end{tabular}
	\egroup
\end{table}

Let us again investigate the performance of the ADDA augmented equivariant model on a class by class basis. Performance of the original classifier was near perfect, achieving AUC scores of $\approx 0.99$ for all three classes (e.g. subhalos vs. vortices \textit{and} no substructure, etc.) on the source data set. Applied naively to the target data set it achieved a respectable performance of $\approx 0.92$ for subhalos but only managed AUC scores of $\approx 0.83$ and $0.82$ for no substructure and vortices respectively. Note that this is actually \textit{worse} than the comparable \textit{ResNet-18} application. ADDA dramatically increases the algorithm's performance, managing consistent classification across all classes with AUC scores of $\approx 0.98$. Considering the naive performance was worse than \textit{ResNet-18}, it is impressive that our equivariant model ENN was augmented more effectively with ADDA, a feature that is shared with the other two UDA methods. 

Now let us consider the performance of the ENN relative to \textit{ResNet-18}. We can see from Figure~\ref{fig:roc} that there was a notable increase in the ability of AdaMatch and Self-Ensemble to improve performance for the ENN. This is likely related to the fact that both AdaMatch and Self-Ensemble are highly dependent on augmentations. Thus, the additional symmetries manifest in the ENN, relative to CNNs, can be thought of as negating the redundancies of augmentation. Concretely, data augmentation is less important for the ENN, since augmented data (rotations, translation, reflections, etc.) are invariant under a  gauge transformation. On the other hand, CNNs only realize translational invariance, which means they are susceptible to idiosyncrasies induced in training.

\section{Discussion \& Conclusion} \label{sec:DNC}

With the upcoming arrival of strong lensing data from Euclid and the Vera Rubin Observatory, it is imperative to assess how algorithms trained on simulations can be adapted to study real world data. To this goal, in this work we studied how unsupervised domain adaptation algorithms can be used to adapt a model trained on one set of data (the source) to another, more complex, data set (the target). To make a precise quantitative evaluation, we based it on two sets of simulations, with the more complex simulation as a proxy for real data. 

We have demonstrated that the naive application of substructure classification models have diminished performance when applied to a more complex target data set. We then tested the implementation of several UDA techniques (ADDA, AdaMatch, and Self-Ensemble) for a popular convolutional (\textit{ResNet-18}) classifier and a symmetry equivariant (ENN) classifier. We found that UDA consistently increases the performance of all models on the target data set. The ENN-based ADDA algorithm achieved top performance, achieving performance competitive with the original source trained and evaluated algorithm. 

Investigating performance on a class by class basis, we found that classification performance is consistent between classes. This is despite the fact that the naive application of the source-trained ENN has significantly degraded AUC scores for no substructure and vortices relative to subhalos. This observation, UDA aside, is encouraging as the base architectures are relatively robust to the signature of subhalos lensing data, a common observable among many dark matter models. This result is not surprising as the lensing signature of vortices is inherently more difficult relative to subhalos since they induce no magnification of the background source (see \cite{Alexander:2019puy} and references therein for details). Nonetheless, it is impressive that UDA can completely compensate for this extra difficulty.

With the upcoming arrival of high quality strong lensing data, domain adaptation techniques will be critical for real world applications of machine learning based dark matter analyses. Possible performance degradations for a simulation trained model naively applied to real data sets can be more significant than what was realized here, making the need for further development and application of UDA methods even more critical. While we have restricted ourselves to substructure classification in this work, domain adaptation techniques can be additionally useful in the broader context of studying dark matter, from regression to image segmentation, in applications to real world strong gravitational lensing data sets.

\section{Acknowledgements}
 M.~T. was a participant in the Google Summer of Code (GSoC) 2021 program. S.~G. was supported in part by the National Science Foundation Award No. 2108645. S.~A. and M.~W.~T. were supported in part by U.S. National Science Foundation Award No. 2108866. This work made use of these additional software packages: \textit{Matplotlib} \cite{Hunter:2007}, \textit{NumPy} \cite{harris2020array}, \textit{PyTorch} \cite{NEURIPS2019_9015}, and \textit{SciPy} \cite{2020SciPy-NMeth}.

\bibliography{bib.bib}

\end{document}